\documentclass[twocolumn,runningheads]{svjour2}
\smartqed  
\usepackage{graphicx}
\usepackage{mathptmx} 
\journalname{Astrophysics and Space Science}
\begin{document}

\title{A microscopic equation of state for protoneutron stars
\thanks{Talk given by G.F. Burgio}
}
\titlerunning{A microscopic equation of state for protoneutron stars}

\author{G.F. Burgio \and M. Baldo \and O.E. Nicotra \and H.-J. Schulze}

\authorrunning{G.F. Burgio et al.}

\institute{
 G.F. Burgio \at
 INFN Sezione di Catania, Via S. Sofia 64, I-95123 Catania, Italy \\
 \email{fiorella.burgio@ct.infn.it}
 \and
 M. Baldo \at
 INFN Sezione di Catania, Via S. Sofia 64, I-95123 Catania, Italy \\  
 \and
 O.E. Nicotra \at
 INFN Sezione di Catania, Via S. Sofia 64, I-95123 Catania, Italy \\  
 \and
 H.-J. Schulze \at
 INFN Sezione di Catania, Via S. Sofia 64, I-95123 Catania, Italy}

\date{Received: date / Accepted: date}

\maketitle

\begin{abstract}
We study the structure of protoneutron stars within the finite-tem\-perature 
Brueckner-Bethe-Gold\-stone ma\-ny-bo\-dy theory. 
If nucleons, hyperons, and leptons are present in the stellar core, 
we find that neutrino trapping stiffens considerably the
equation of state, because hyperon onsets are shifted to larger 
baryon density. 
However, the value of the critical mass turns out to be smaller than 
the ``canonical'' value 1.44 $M_\odot$. 
We find that the inclusion of a hadron-quark phase transition 
increases the critical mass and stabilizes it at about 1.5--1.6 $M_\odot$.
\end{abstract}

\section{Introduction}
\label{intro} 

After a protoneutron star (PNS) is successfully formed
in a supernova explosion, neutrinos are temporarily trapped wi\-thin
the star (Prakash et al. 1997). The subsequent evolution of the PNS
is strongly dependent on the stellar composition, which is mainly
determined by the number of trapped neutrinos, and by thermal
effects with values of temperatures up to 30--40 MeV 
(Burrows and Lattimer 1986; Pons et al. 1999). 
Hence, the equation of state (EOS)
of dense matter at finite temperature is crucial for studying the
macrophysical evolution of PNS.

Only a few microscopic calculations of the nuclear EOS at finite
temperature are available so far. The variational calculation by
Friedman and Pandharipande (1981) was one of the first
semi-microscopic investigations of the finite-tem\-pe\-rature EOS. The
results predict a Van der Waals behavior for symmetric matter, 
which leads to a
liquid-gas phase transition with a critical temperature 
\hbox{$T_c \approx$ 18--20 MeV.} 
Later, Brueckner calculations 
(Lejeune et al. 1986; Baldo and Ferreira 1999) 
and chiral perturbation theory at
finite temperature (Kaiser et al. 2002) confirmed these findings
with very similar values of $T_c$. The Van der Waals behavior was
also found in the finite-temperature relativistic Dirac-Brueckner
calculations of Ter Haar and Malfliet (1986, 1987) and Huber et al.
(1999), although at a lower temperature.

We have developed a microscopic EOS in the framework of the
Brueckner-Bethe-Goldstone (BBG) many-body approach 
including nucleons and hyperons and extended to finite temperature. 
This EOS has been successfully applied to the
study of the limiting temperature in nuclei 
(Baldo et al. 1999, 2004). 
Recently, this EOS has been extended for
including neutrino trapping, and results have been presented in
Nicotra et al. (2006a). 

The scope of this work is to discuss
composition and structure of these newly born stars with the EOS
previously mentioned, also including a possible transition to quark
matter. In fact, superdense matter in PNS cores may
consist of weakly interacting quarks rather than hadrons, due to the
asymptotic freedom. The appearance of quarks can alter substantially
the chemical composition of a PNS, with observable
consequences on the evolution, like onset of metastability and
abrupt cessation of the neutrino signal (Pons et al. 2001).

We have studied the effects of a hadron-quark phase transition within
the MIT bag model, and found that the presence of quark matter increases 
the value of the maximum mass of a PNS, and
stabilizes it at about 1.5-1.6 $M_\odot$, 
no matter the value of the temperature.

This paper is organized as follows. In Sec.~2 we briefly review the 
BBG theory of nuclear matter at finite temperature, 
including both nucleons and hyperons. In Sec.~3 we discuss the chemical 
composition of a PNS, whereas in Sec.~4 we study its structure, 
even including a possible transition to a quark phase. Finally, in Sec.~5, 
we draw our conclusions.

\section{The BBG EOS at finite temperature}
\label{sec:1}

In the recent years, the BBG perturbative theory has made much
progress, since its convergence has been firmly established 
(Day 1981; Song et al. 1998; Baldo et al. 2000b; Sartor 2006), 
and has been extended in a fully
microscopic and self-consistent way to the hyperonic sector 
(Schulze et al. 1995, 1998, 2006; Baldo et al. 1998, 2000a). 
Moreover, the addition
of phenomenological three-body forces (TBF) based on the Urbana
model (Carlson et al. 1983; Schiavilla et al. 1986), permitted to
improve to a large extent the agreement with the empirical
saturation properties (Baldo et al. 1997; Zhou et al. 2004).

The finite-temperature formalism which is closest to the BBG expansion, 
and actually reduces to it in the zero-tem\-pe\-ra\-ture limit, 
is the one formulated by Bloch and De Dominicis (1958, 1959). 
In this approach the
essential ingredient is the two-body scattering matrix $K$, which,
along with the single-particle potential $U$, satisfies the
self-consistent equations
\begin{eqnarray}
  \langle k_1 k_2 | K(W) | k_3 k_4 \rangle
 &=& \langle k_1 k_2 | V | k_3 k_4 \rangle
\nonumber\\&&
 \hskip -30mm +\; \mathrm{Re}\!\!\sum_{k_3' k_4'}
 \langle k_1 k_2 | V | k_3' k_4' \rangle\!
 { [1\!-n(k_3')] [1\!-n(k_4')] \over
   W - E_{k_3'} - E_{k_4'} + i\epsilon }
 \langle k_3' k_4' | K\!(\!W\!) | k_3 k_4 \rangle
\nonumber\\&& 
\label{eq:kkk}
\end{eqnarray}
and
\begin{equation}
 U(k_1) = \sum_{k_2} n(k_2) \langle k_1 k_2 | K(W) | k_1 k_2 \rangle_A \:,
\label{eq:ueq}
\end{equation}
where $k_i$ generally denote momentum, spin, and isospin. Here $V$
is the two-body interaction, and we choose the Argonne $V_{18}$
nucleon-nucleon potential (Wiringa et al. 1995).
$W = E_{k_1} + E_{k_2}$ represents the
starting energy, and $E_k = k^2\!/2m + U(k)$ the single-particle
energy. Eq.~(\ref{eq:kkk}) coincides with the Brueckner equation for
the $K$ matrix at zero temperature, if the single-particle
occupation numbers $n(k)$ are taken at $T = 0$. At finite
temperature $n(k)$ is a Fermi distribution. For a given density and
temperature, Eqs.~(\ref{eq:kkk}) and (\ref{eq:ueq}) have to be
solved self-consistently along with the equation for the
grand-canonical potential density $\omega$ and the free energy
density, which has the following simplified expression
\begin{equation}
 f = \sum_i \left[ \sum_{k} n_i(k)
 \left( {k^2\over 2m_i} + {1\over 2}U_i(k) \right) - Ts_i \right] \:,
 \label{fr_en}
\end{equation}
where
\begin{equation}
 s_i = - \sum_{k} \Big( n_i(k) \ln n_i(k) + [1-n_i(k)] \ln [1-n_i(k)] \Big)
\end{equation}
is the entropy density for component $i$ treated as a free gas with
spectrum $E_i(k)$. For a more extensive discussion of this topic,
the reader is referred to (Baldo 1999) and references therein.

In deriving Eq.~(\ref{fr_en}), we have introduced the so-called
{\em Frozen Correlations Approximation}, 
i.e., the correlations at $T\neq 0$ are assumed to be essentially 
the same as at $T=0$. 
This means that the single-particle potential $U_i(k)$ for the component $i$ 
can be approximated by the one calculated at $T=0$. 
This allows to save computational time and simplify the numerical procedure. 
It turns out that the assumed independence is
valid to a good accuracy, at least for not too high temperature
(Baldo and Ferreira 1999, Fig.~12).

\begin{figure}[t] 
\centering
\includegraphics[width=7.5cm,angle=270]{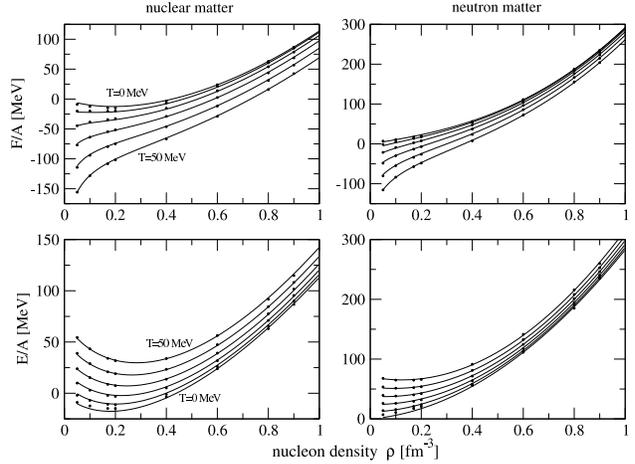}
\caption{
Finite-temperature BBG EOS for symmetric (left-hand panels) and
purely neutron (right-hand panels) matter.
The upper panels show the free energy and the lower panels the
internal energy per particle as a function of the nucleon density.
The temperatures vary from 0 to 50 MeV in steps of 10 MeV.}
\label{f:ba}
\end{figure} 

In our many-body approach, we have also introduced TBF
among nucleons, in order to reproduce correctly the nuclear
matter saturation point $\rho_0 \approx 0.17~\mathrm{fm}^{-3}$, $E/A
\approx -16$ MeV. Since a complete microscopic theory of TBF is not
available yet, we have adopted the phenomenological Urbana model
(Carlson et al. 1983; Schiavilla et al. 1986), which consists of an
attractive term due to two-pion exchange with excitation of an
intermediate $\Delta$ resonance, and a repulsive phenomenological
central term. In the BBG approach, the TBF is reduced to a 
density-de\-pen\-dent 
two-body force by averaging over the position of the third
particle, assuming that the probability of having two particles at a
given distance is reduced according to the two-body correlation
function. The corresponding EOS reproduces correctly the nuclear
matter saturation point (Baldo et al. 1997; Zhou et al. 2004), and
gives values of incompressibility and symmetry energy at saturation
compatible with those extracted from phenomenology 
(Myers and Swiatecki 1996).

\begin{figure*}[t] 
\centering
\includegraphics[height=12.5cm,angle=270]{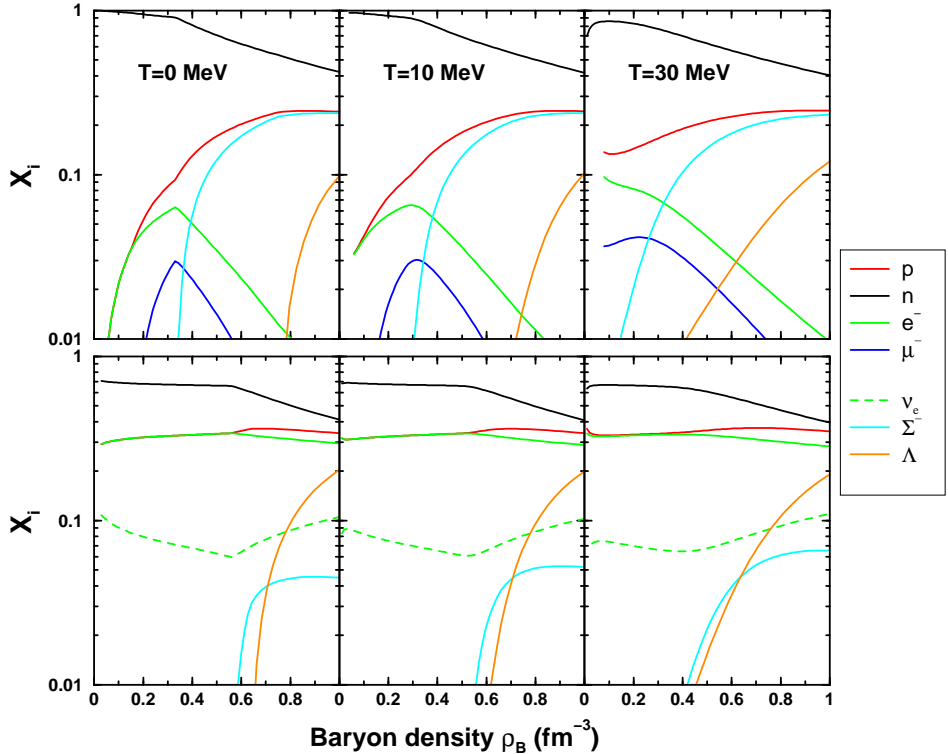} 
\caption{
Relative populations for neutrino-free (upper panels) and 
neutrino-trapped (lower panels) matter as a
function of the baryon density for several values of the
temperature.} \label{f:xi}
\end{figure*} 

In Fig.~\ref{f:ba} we display the free energy (upper panels) and the
internal energy (lower panels) obtained following the above
discussed procedure, both for symmetric and purely neutron matter,
as a function of the nucleon density. Calculations are reported for
several values of the temperature between 0 and 50 MeV. We notice
that the free energy of symmetric matter shows a typical Van der
Waals behavior (with $T_c\approx 16$ MeV) and is a monotonically
decreasing function of the temperature. On the contrary, the
internal energy is an increasing function of the temperature. At
$T=0$ the free energy coincides with the total energy and the
corresponding curve is just the usual nuclear matter saturation
curve.

\subsection{Inclusion of hyperons}

The fast rise of the baryon chemical potentials with density in
neutron star cores (Glendenning 1982, 1985) may trigger the
appearance of strange baryonic species, i.e., hyperons.  For this
purpose we have extended the BBG approach in order to include the
$\Sigma^-$ and $\Lambda$ hyperons 
(Schulze et al. 1998, 2006; Baldo et al. 1998, 2000a). 
The inclusion of hyperons requires the knowledge of
the nucleon-hyperon (NH) and hyperon-hyperon (HH) interactions. In
our past papers, we have shown results obtained at $T=0$, and we have
used the Nijmegen soft-core NH potential (Maessen et al. 1989), and
neglected the HH interactions, since no reliable HH potentials are
available yet. For these reasons, we present in this article
finite-temperature calculations using free hyperons. A more complete
set of calculations obtained with the inclusion of the NH
interaction at finite temperature will be published elsewhere.

We have found that due to its negative charge the $\Sigma^-$ hyperon
is the first strange baryon appearing in the reaction $n+n
\rightarrow p+\Sigma^-$, in spite of its substantially larger mass
compared to the neutral $\Lambda$ hyperon
($M_{\Sigma^-}=1197\;\mathrm{MeV}, M_\Lambda=1116\;\mathrm{MeV}$). The
presence of hyperons strongly softens the EOS, mainly due to the
larger number of baryonic degrees of freedom. 
This EOS produces a
maximum neutron star mass that lies slightly below the canonical
value of 1.44 $M_\odot$ (Taylor and Weisberg 1989),
as confirmed by Schulze et al. (2006)
also with the NSC97 hyperon potentials (Stoks and Rijken 1999).
This could indicate the presence of non-baryonic (quark) matter in the interior
of heavy neutron stars (Burgio et al. 2002a, 2002b; Baldo et al.
2003; Maieron et al. 2004). 
This point is discussed more extensively below.

\section{Composition and EOS of hot stellar matter}

For stars in which the strongly interacting particles are only baryons,
the composition is determined by the requirements of charge
neutrality and equilibrium under the weak processes
\begin{equation}
 B_1 \rightarrow B_2 + l + {\overline \nu}_l \ ,\quad
 B_2 + l \rightarrow B_1 + \nu_l \:,
\label{weak:eps}
\end{equation}
where $B_1$ and $B_2$ are baryons and $l$ is a lepton,
either an electron or a muon.
When the neutrinos are trapped, these two requirements imply that
the relations
\begin{equation}
 \sum_i q_i x_i + \sum_l q_l x_l  = 0
\label{neutral:eps}
\end{equation}
and
\begin{equation}
\mu_i = b_i \mu_n - q_i( \mu_l - \mu_{\nu_l} )
\label{nitrap:eps}
\end{equation}
are satisfied.
In the expression above, $x_i=\rho_i/\rho_B$ represents the
baryon fraction for the species $i$, 
$\mu_i$ the chemical potential, $b_i$ the baryon number, 
and $q_i$ the electric charge. 
Equivalent quantities are defined for the leptons $l$.

For stellar matter containing nucleons and hyperons as relevant degrees of
freedom, the chemical equilibrium conditions read explicitly
\begin{eqnarray}
 \mu_n - \mu_p &=& \mu_e - \mu_{\nu_e} = \mu_\mu + \mu_{\bar{\nu}_\mu} \:,
\nonumber\\
\mu_{\Sigma^-} &=& 2 \mu_n - \mu_p \:,
\nonumber\\
  \mu_\Lambda  &=& \mu_n \:. \label{beta:eps}
\end{eqnarray}
The nucleon chemical potentials are calculated starting from the free
energy and its partial derivatives with respect to the total baryon
density $\rho$ and proton fraction $x_p$, i.e.,
\begin{eqnarray}
 \mu_{n}(\rho,x_{p}) &=&
 \left[ 1 + \rho \frac{\partial }{\partial\rho}
        -x_{p}\frac{\partial }{\partial x_{p}} \right] {f\over\rho} \:,
\label{mun:eps}
\\
 \mu_{p}(\rho,x_{p}) &=&
  \left[ 1 + \rho\frac{\partial}{\partial\rho}
         + (1-x_p)\frac{\partial}{\partial x_p} \right] {f\over\rho} \:,
\label{mup:eps}
\end{eqnarray}
whereas 
the chemical potentials of the noninteracting leptons and hyperons
are obtained by solving numerically the free
Fermi gas model at finite temperature.
More details are given in Nicotra et al. (2006a).

Because of trapping, the numbers of leptons per baryon
of each flavor $l=e,\mu$,
\begin{equation}
 Y_l = x_l - x_{\bar l} + x_{\nu_l} - x_{\bar{\nu}_l} \:,
\label{lepfrac:eps}
\end{equation}
are conserved on dynamical time scales.
Gravitational collapse
calculations of the white-dwarf core of massive stars indicate that at
the onset of trapping, the electron lepton number
$Y_e = x_e + x_{\nu_e}\approx 0.4$,
the precise value depending on the efficiency of electron capture
reactions during the initial collapse stage.
Moreover, since no muons are present when neutrinos become trapped,
the constraint
$Y_\mu = x_\mu - x_{\bar{\nu}_\mu} = 0$
can be imposed.
We fix the $Y_l$ at these values in our calculations
for neutrino-trapped matter.

Let us now discuss first the populations of beta-equi\-li\-brated
stellar matter, by solving the chemical equilibrium conditions given
by Eqs.~(\ref{beta:eps}), supplemented by electrical char\-ge
neutrality and baryon number conservation. In Fig.~\ref{f:xi} we
show the particle fractions as a function of baryon density, for
different values of the temperature. The upper panels show the particle
fractions when stellar
matter does not contain neutrinos, whereas the lower panels
show the populations in neutrino-trapped matter.
We observe that the electron fraction is larger in neutrino-trapped 
than in neutrino-free matter, and, as a consequence, the proton population 
is larger.

Neutrino trapping strongly influences the onset of
hyperons. In fact, the threshold density of the $\Sigma^-$ is shifted to high
density, whereas $\Lambda$'s appear at slightly smaller density.
This is due to the fact that the $\Sigma^-$ onset depends on the
neutron and lepton chemical potentials, i.e., $\mu_n + \mu_e -
\mu_{\nu_e}$, which stays at larger values in neutrino-trapped
matter than in the neutrino-free case 
because of the larger fraction of electrons, 
thus delaying the appearance
of the $\Sigma^-$ to higher baryon density and limiting its
population to a few percent. On the other hand, the $\Lambda$ onset
depends on the neutron chemical potential only, which keeps at lower
values in the neutrino-trapped case. When the temperature increases, 
more and more hyperons are present also at low densities, but they represent 
only a small fraction
of the total baryon density in this region of the PNS.
Altogether, the hyperon fractions are much smaller than in the
neutrino-free matter. Therefore the corresponding EOS will be stiffer
than in the neutrino-free case.

\begin{figure}[t] 
\centering
\includegraphics[width=9.4cm,angle=270]{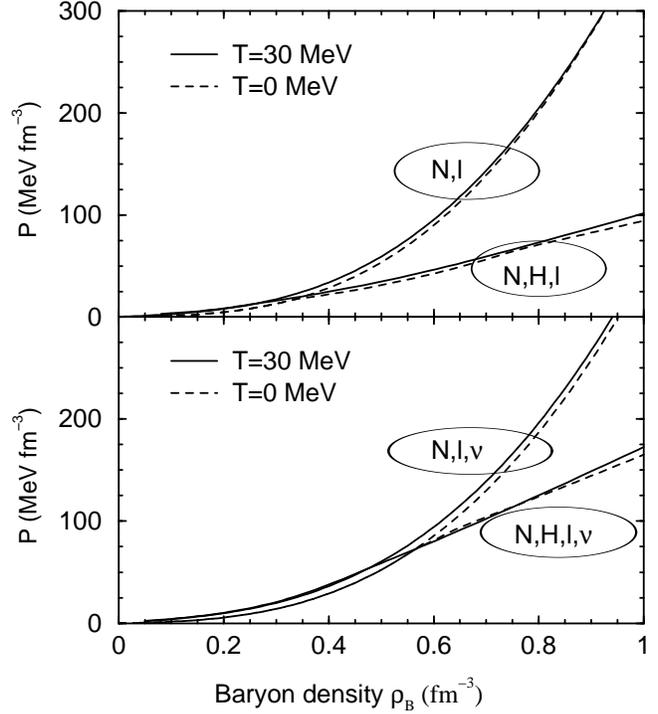} 
\caption{
Pressure as a function of baryon density for
beta-equilibrated matter at temperatures $T=0$ and 30 MeV.
The upper (lower) panel shows the EOS in
neutrino-free (neutrino-trapped) matter,
with nucleons only (upper curves) and nucleons plus hyperons (lower curves).}
\label{f:eos}
\end{figure} 

Once the composition of the $\beta$-stable stellar matter is known,
one can proceed to calculate the free energy density $f$ and then the 
pressure $p$ through the usual thermodynamical relation
\begin{equation}
 p = \rho^2 {\partial{(f/\rho)}\over \partial{\rho}} \:.
\end{equation}
The resulting EOS is displayed in Fig.~\ref{f:eos}, where the
pressure for beta-stable asymmetric matter, without (upper panel) 
and with (lower panel) neutrinos, is plotted as a function of the 
baryon density at temperatures $T=0$ and 30 MeV. 
Let us begin with discussing the
case of neutrino-free matter, shown without (upper curves) and
with hyperons (lower curves). We notice that thermal effects
produce a slightly stiffer EOS with respect to the cold case, and
that at very high densities they almost play no role. 

In the case that only nucleons are present, the EOS gets softer with 
increasing temperature at high baryon density. This behavior is at variance 
with the results obtained by Prakash et al. (1997). In this regard, we 
should notice that in our calculations we are considering an isothermal 
profile, whereas in Prakash et al. (1997) the profile is isentropic.
Another difference between the two approaches is in the many-body method.
A complete comparison can be made only by adopting an isentropic description 
within our BHF approach. This matter is left to further investigations.

The inclusion of hyperons produces a dramatic effect, because the EOS gets much
softer, no matter the value of the temperature.
In this case, thermal effects dominate over the whole density range 
since on the average the Fermi energies are smaller. A similar behavior was 
found in Prakash et al. (1997).

In the lower panel we show the corresponding neutrino-trapped
case. The EOS is slightly softer than in the neutrino-free case if
only nucleons and leptons are present in the stellar matter. Again,
the presence of hyperons introduces a strong softening of the EOS,
but less than in the neutrino-free case, because now the hyperons
appear later in the matter and their concentration is lower. 
Thermal effects are rather
small also in this case, except for the disappearance of the hyperon
onsets.

\section{(Proto)neutron star structure}

The stable configurations of a (proto)neutron star can be obtained
from the well-known hydrostatic equilibrium equations of Tolman,
Oppenheimer, and Volkov (Shapiro and Teukolsky 1983) for the pressure
$p$ and the enclosed mass $m$,
\begin{eqnarray}
 {dp(r)\over dr} &=& -\frac{Gm(r)\epsilon(r)}{r^2}
 \frac{\big[ 1 + {p(r)\over\epsilon(r)} \big]
       \big[ 1 + {4\pi r^3p(r)\over m(r)} \big]}
 {1-{2Gm(r)\over r}} \:,
\label{tov1:eps}
\\
 \frac{dm(r)}{dr} &=& 4\pi r^{2}\epsilon(r) \:,
\label{tov2:eps}
\end{eqnarray}
once the EOS $p(\epsilon)$ is specified, being $\epsilon$ the total
energy density ($G$ is the gravitational constant). For a chosen
central value of the energy density, the numerical integration of
Eqs.~(\ref{tov1:eps}) and (\ref{tov2:eps}) provides the mass-radius
relation.

For the description of the (proto)neutron star crust, we have joined
the hadronic EOS described above with the ones by
Negele and Vautherin (1973) in the medium-density regime
($0.001\;\mathrm{fm}^{-3}<\rho<0.08\;\mathrm{fm}^{-3}$), and the
ones by Feynman, Metropolis, and Teller (1949) and Baym, Pe\-thi\-ck, and
Sutherland (1971) for the outer crust 
($\rho<0.001\;\mathrm{fm}^{-3}$).

\begin{figure}[t] 
\centering
\includegraphics[width=10.6cm,angle=270]{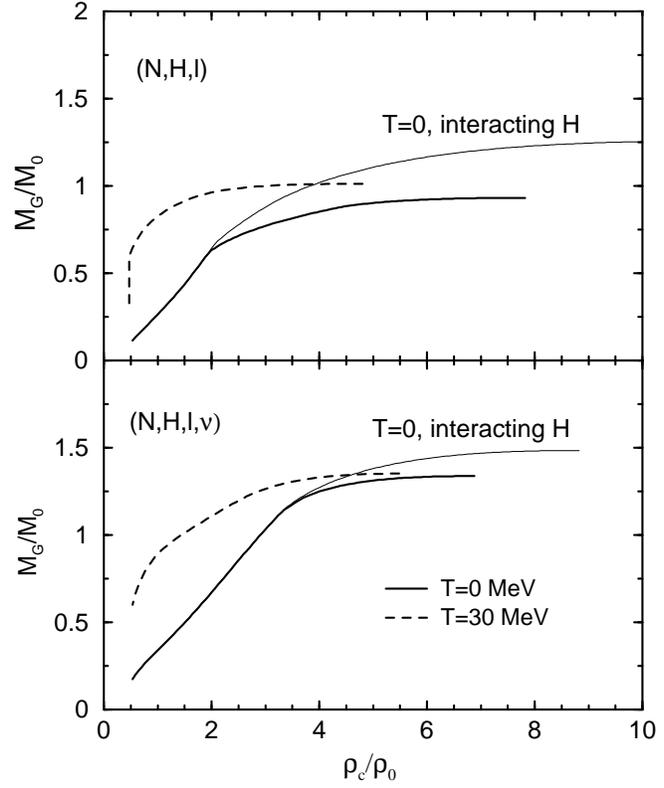} 
\caption{
The gravitational mass
(in units of the solar mass)
as a function of the central baryon density
(normalized with respect to the saturation value
$\rho_0=0.17\;\mathrm{fm}^{-3}$)
at temperatures $T=0$ and 30 MeV.
The upper (lower) plot regards neutrino-free (neutrino-trapped)
matter. The thin solid curves denote configurations of
cold stars employing interacting hyperons, whereas the remaining curves show
results obtained using free hyperons.}
\label{f:mg}
\end{figure} 

Simulations of supernovae explosions (Burrows and Lattimer 1986; Pons
1999) show that the PNS has neither an isentropic nor
an isothermal profile. For simplicity we will assume a constant
temperature inside the star and attach a cold crust for the outer
part. This schematizes the temperature profile of the PNS.
More realistic temperature profiles can be obtained by
modelling the neutrinosphere both in the interior and in the
external outer envelope, which is expected to be much cooler. A
proper treatment of the transition from the hot interior to the cold
outer part can have a dramatic influence on the mass-central density
relation in the region of low central density and low stellar
masses. In particular, the ``minimal mass" region, typical of cold
neutron stars (Zel'dovich and Novikov 1971; Shapiro and Teukolsky 1983), 
can be shifted in PNS to much higher values of central density and
masses. A detailed analysis of this point can be found in (Gondek et
al. 1997), where a model of the transition region between the
interior and the external envelope is developed. However, as
discussed in the next section, the maximum mass region is not
affected by the structure of this low-density transition region.

In Fig.~\ref{f:mg} we show the gravitational mass
(in units of the solar mass $M_\odot=1.98\times 10^{33}$g)
as a function of the central baryon density
for stars containing hyperons. 
We observed in Fig.~\ref{f:eos} that the EOS softens considerably when
hyperons are included, both in neutrino-free and neutrino-trapped matter.
As a consequence the mass -- central density relation is also significantly
altered by the presence of hyperons and the value of the critical mass 
is about $1.3\;M_\odot$ for neutron stars (upper plot, thin curve)
and $1.5\;M_\odot$
for protostars in the $\nu$-trapped stage (lower plot, thin curve). 
Those results are obtained employing interacting hyperons
at zero temperature, choosing the Njimegen potential as 
nucleon-hyperon potential (Maessen et al. 1989), which is well adapted to the 
available experimental NH scattering data.   
However, since the value of the critical mass for cold stars falls below 
the mass of the best observed pulsar, i.e., $1.44~M_\odot$ 
(Taylor and Weisberg 1989),
the EOS of high-density nuclear matter comprising only baryons
(nucleons and hyperons) is probably unrealistic
(even taking into account the present uncertainty
of hyperonic two-body and three-body forces),
and must be supplemented by a transition to quark matter.
This has been discussed extensively in
(Burgio et al. 2002a, 2002b; Baldo et al. 2003; Maieron et al. 2004), and 
will be briefly recalled in the next subsection.

\begin{figure}[t] 
\centering
\includegraphics[width=7.2cm,angle=270]{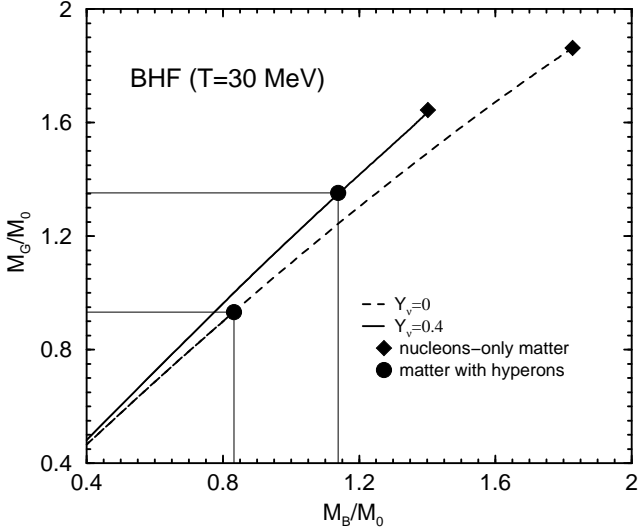} 
\caption{
Gravitational mass as a function of the baryon mass
for neutrino-trapped matter at temperature $T=30$ MeV (solid curve) 
and for cold neutrino-free matter (dashed curve).
A dot at the end of the curves indicates matter with 
noninteracting hyperons, 
a diamond indicates purely nucleonic matter.}
\label{f:mb}
\end{figure} 

Nevertheless, it is interesting to observe that the maximum mass of hyperonic 
protostars is larger (by about $0.3\;M_\odot$)
than the one of cold neutron stars.
The reason is the minor importance of hyperons in the neutrino-saturated 
matter, which leads to a stiffer EOS, see Fig.~\ref{f:eos},
and to a larger maximum mass. This feature could lead to metastable stars 
suffering a delayed collapse while cooling down,
as discussed in (Prakash et al. 1997; Pons et al. 1999).
Metastable stars occur within a range of masses near the maximum mass 
of the initial configuration and remain stable only for several seconds after 
formation. In order to study metastability, it is useful to calculate 
the baryonic mass $M_B$, which is proportional to the number of baryons 
in the system and is constant during the evolution of the isolated star,
if no accretion is assumed during the entire PNS evolution.
For that, Eqs.~(\ref{tov1:eps}) and (\ref{tov2:eps}) must be supplemented with
\begin{equation}
 {dM_B(r)\over dr} = \frac{4\pi r^2 \rho_B m_N}{\sqrt{1-2Gm(r)/r}} \:,
\label{tov3:eps}
\end{equation}
where $m_N= 1.67\times 10^{-24}$g is the nucleon mass.
As one can see from Fig.~\ref{f:mb}, if hyperons are present 
(lines ending with a dot), then deleptonization, 
that is the transition from $Y_\nu=0.4$ to $Y_\nu=0$, 
lowers the range of baryonic masses that can be 
supported by the EOS from about 1.15 $M_\odot$ to about
0.84 $M_\odot$. The window in the baryonic mass in which neutron stars are 
metastable is thus about 0.31 $M_\odot$. On the other hand, if hyperons 
are absent (lines ending with a diamond), the maximum baryonic mass increases 
during deleptonization, and no metastability occurs.

\subsection{Including quark matter}
\label{subsec:4}

The appearance of quark matter in the interior of massive neutron stars
is one of the main issues in the physics of these compact objects.
Calculations of neutron star structure, based on microscopic nucleonic
EOS, indicate that the central particle density
may reach values larger than $1/\rm fm^{3}$.
In this density range the nucleon cores (dimension $\approx 0.5\;\rm fm$)
start to touch each other, and it is hard to imagine that only
nucleonic degrees of freedom can play a role.
Rather, it can be expected that even before reaching
these density values, the nucleons start to loose their identity, and quark
degrees of freedom are excited at a macroscopic level.
Unfortunately, while the microscopic theory of the nucleonic EOS has
reached a high degree of sophistication, the quark matter EOS is poorly
known at zero temperature and at the high baryonic density
appropriate for NS.

\begin{figure}[t] 
\centering
\includegraphics[width=7.0cm,angle=270]{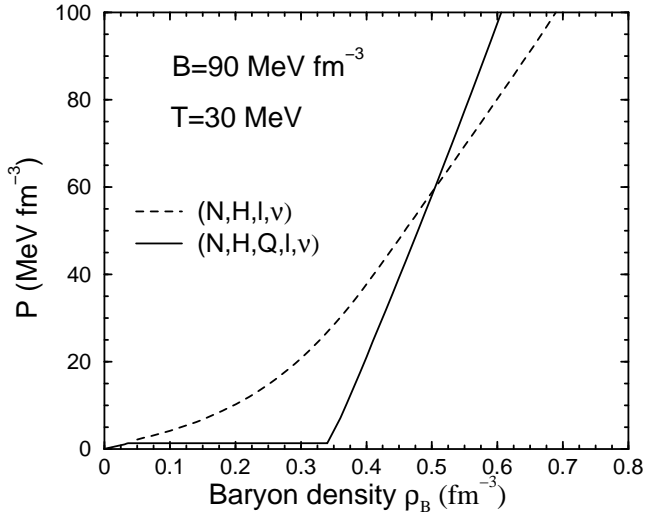}
\caption{
Pressure as a function of the baryon density 
at temperature $T=30\;\rm MeV$, 
and a bag constant $B=90\;\rm MeVfm^{-3}$.
The solid (dashed) curve is for neutrino-trapped matter 
containing baryons and quark matter (only baryons).}
\label{f:eosqm}
\end{figure} 

Here we will briefly discuss some results obtained by using the MIT bag 
model (Chodos et al. 1974) at finite temperature, 
with a bag constant $B=90\;\rm MeVfm^{-3}$. 
In order to study the hadron-quark phase transition, we have performed a
Maxwell construction between the baryon phase and the quark phase. 
A more realistic model is the Glendenning construction (Glendenning 1992), 
which determines the 
range of baryon density where both phases coexist, yielding an EOS containing
a pure hadronic phase, a mixed phase, and a pure quark matter region. 
Using the Maxwell construction implies that the phase transition is sharp, and 
no mixed phase exists. The pressure is characterized by a plateau extended
over a range of baryon density, which separates the purely hadron phase at 
low density from the pure quark phase at higher density. 

This is shown in 
Fig.~\ref{f:eosqm}, which displays the case of hot, neutrino-trapped matter.
The solid line represents the EOS with the hadron-quark phase transition, 
whereas the dashed line is the purely hadronic case. As we can see, the EOS 
which comprises the hadron and the quark phases is stiffer at large baryon 
density than the purely hadronic case. A stiffer EOS causes 
larger values of the critical mass for a neutron star, as we found in 
(Burgio et al. 2002b).

In particular, the critical mass may reach values up to 
1.5--1.6 $M_\odot$, depending on the value of the bag constant. The
maximum mass increases with decreasing value of $B$, but the latter
cannot be too small if stability of symmetric nuclear matter at
saturation has to be ensured. We have also checked that these
results are quite general, and do not change appreciably if a
density-dependent bag constant is introduced (Nicotra et al. 2006b).

In Table 1, we show the values of the critical mass for PNS
with trapped neutrinos at temperatures T=30 and 50 MeV 
and compare with the value for a cold NS.
We find that the hadron-quark phase 
transition stabilizes the value of the maximum mass, which turns out to be 
independent of the temperature and equal to 1.53 $M_\odot$ for the adopted 
value of the bag constant. 
In this case we do not find any metastable
stars, at variance with the findings of Prakash et al. (1997).
More detailed 
calculations will be presented elsewhere (Nicotra et al. 2006b).

\section{Conclusions}

\begin{table}[t] 
\caption{
The values of the maximum mass 
and corresponding central density
(normalized with respect to the saturation density $\rho_0=0.17\;\rm fm^{-3}$)
for different stellar configurations.}
\centering
\label{tab:1}
\begin{tabular}{r|lll}
\hline\noalign{\smallskip}
    &  $T$(MeV) & $M_G/M_\odot$ & $\rho_c/\rho_0$ \\[3pt]
\tableheadseprule\noalign{\smallskip}
PNS & 30 & 1.53 & 8.2 \\
PNS & 50 & 1.53 & 7.9 \\
 NS &  0 & 1.5 & 9.5 \\
\noalign{\smallskip}\hline
\end{tabular}
\end{table} 

In this paper we have studied the structure of (proto)neu\-tron stars
on the basis of a microscopically derived EOS for baryonic matter
at finite temperature, in the framework of the Brueckner-Hartree-Fock 
many-body theory.
Configurations with or without trapped neutrinos were considered.
We found that the maximum mass of a hyperonic
PNS is substantially larger than the one of the cold star,
because both neutrino trapping and finite temperature tend to
stiffen the EOS. Trapping shifts the onset of hyperons,
in particular the $\Sigma^-$, 
to considerably higher density and reduces their concentrations.

However, as in the case of cold neutron stars, the addition of hyperons
demands for the inclusion of quark degrees of freedom in order to obtain
a maximum mass larger than the observational lower limit.
For this purpose, we have studied the hadron-quark phase transition within the 
MIT bag model, and performed a Maxwell construction between the two phases.
We found that the inclusion of quark matter stabilizes the value of the 
critical mass of a PNS at about 1.5--1.6 $M_\odot$,
no matter the value of the temperature.


\end{document}